\documentclass[10pt,onecolumn]{myart}  
\usepackage{graphics}
\usepackage{amssymb}
\newcommand{\vek}[1]{\mbox{\bf #1}}

\begin{document}
\begin{frontmatter}

\title{Hamiltonians and Green's functions which interpolate
between two and three dimensions}

\author{Rainer Dick\thanksref{rainer}}

\address{Department of Physics and Engineering Physics,
        University of Saskatchewan,\\ 116 Science Place, Saskatoon, 
        SK S7N 5E2, Canada}

\thanks[rainer]{rainer@sask.usask.ca}

\begin{abstract}
I propose to use Hamiltonians which contain two-dimensional
and three-dimensional terms for the description of 
two-dimensional systems in physics.

As a model system the evolution of three-dimensional
wavefunctions in the presence of an infinitely thin layer
is studied.
The model predicts distance laws for correlation functions
which interpolate between two-dimensional and three-dimensional
behavior. It also predicts
that in certain cases transmission probabilities
through thin layers should depend not only on the transverse,
but also on the longitudinal
momentum of the infalling particles.

The model also yields a static potential which interpolates between
the two-dimensional logarithmic potential at small distances
and the three-dimensional $1/r$-potential at large distances.
\end{abstract}
\end{frontmatter}

\section{Introduction}\label{intro}

Two dimensions played a prominent role 
in the development of physics in the last twenty years.

On the experimental side this was driven e.g.\ by the needs
of very large scale integration, by
applications of semiconducting layer structures, 
by exploitations of surface catalytic effects,
and by the
development of atomic-scale surface analysis and manufacturing tools
like scanning tunneling microscopy and atomic force microscopy,
to mention only a few developments in this area.

On the theoretical side interest in two-dimensional field theories 
was largely driven
by string theory \cite{GSW}, where fundamental excitations
are described by covariant two-dimensional field theories,
and by the realization that in two-dimensional critical systems
with a rotational symmetry scaling symmetry may be elevated
to full conformal invariance \cite{BPZ}.

The purpose of the present paper is to point out that recent developments
in the mathematical formulation of brane world models may also
inspire new developments in the physics
of low-dimensional systems, and help us to acquire a better understanding
of the transition between three-dimensional and two-dimensional behavior
in these systems. 

The present work was specifically motivated by the
brane world model of Dvali, Ga\-ba\-da\-dze and Porrati, who
recently proposed and analyzed a model which combined gravity on
a $(3+1)$-dimensional manifold (a "3-brane") with gravity in an ambient
$(4+1)$-dimensional bulk \cite{DGP}. 
They observed that the combination
of gravity in different dimensions yields a gravitational potential
which interpolates continuously between the three-dimensional $-1/r$ potential
at small distances and the four-dimensional $-1/r^2$ potential
at large distances, with a transition scale $\ell_{DGP}\simeq m_3^2/m_4^3$
set by the ratios of the reduced Planck masses on the brane and in the bulk.
Since we are not concerned with gravity in the present
paper I will not explicitly write down the model in terms of intrinsic
and extrinsic curvature terms, see \cite{rd1}, but instead refer to
the related model of Dvali, Gabadadze and Shifman \cite{DGS}, which combines 
a Maxwell term in Minkowski space (coordinates $x^\mu=\{t,\vek{r}\}$) 
with a Maxwell term in an ambient
$(4+1)$-dimensional bulk (coordinates $x^M=\{t,\vek{r},x^\perp\}$):
\begin{equation}\label{eq:actionDGS}
S=-\left. \frac{1}{4q_3^2}
\int\! dt\int\! d^{3}\vek{r}\,F_{\mu\nu}F^{\mu\nu}\right|_{x^\perp=0}
-\frac{1}{4q_4^2}\int\! dt\int\! d^{3}\vek{r}\int\! dx^\perp\,
F_{MN}F^{MN}.
\end{equation}
The resulting Coulomb potential on the $(3+1)$-dimensional
Minkowski space\footnote{See Ref.\ \cite{rd2} for a discussion
of dynamical potentials on the brane and in the bulk.} is \cite{DGS}
\begin{equation}\label{eq:Vdgs}
A^0(\vek{r})=\frac{q_3}{4\pi r}\left[
\cos\!\left(\frac{2q_3^2}{q_4^2}r\right)
-
\frac{2}{\pi}\cos\!\left(\frac{2q_3^2}{q_4^2}r\right)
\mbox{Si}\!\left(\frac{2q_3^2}{q_4^2}r\right)
+\frac{2}{\pi}\sin\!\left(\frac{2q_3^2}{q_4^2}r\right)
\mbox{ci}\!\left(\frac{2q_3^2}{q_4^2}r\right)
\right],
\end{equation}
with the sine and cosine integrals
\[
\mbox{Si}(x)=\int_0^x\! d\xi\,\frac{\sin\xi}{\xi},
\qquad
\mbox{ci}(x)=-\int_x^\infty\! d\xi\,\frac{\cos\xi}{\xi}.
\]
The ratio of gauge couplings defines a length scale
\begin{equation}\label{eq:ldgs}
\ell=\frac{q_4^2}{2q_3^2},
\end{equation}
and $A^0$ interpolates between a three-dimensional
distance law at short distances and
a four-dimensional distance law at large distances:
\[
r\ll\ell:
\quad
A^0(\vek{r})=\frac{q_3}{4\pi r}
\left[1+
\frac{2r}{\pi\ell}\!\left(\gamma-1+\ln\!\left(\frac{r}{\ell}\right)\right)
+\mathcal{O}\!\left(\frac{r^2}{\ell^2}\right)\right],
\]
\[
r\gg\ell:
\quad
A^0(\vek{r})=\frac{q_3\ell}{2\pi^2 r^2}
\left[1-2\frac{\ell^2}{r^2}
+ \mathcal{O}\!\left(\frac{\ell^4}{r^4}\right)\right].
\]

Action principles of the kind (\ref{eq:actionDGS}) were denoted as 
dimensionally hybrid action principles in \cite{rd2}.

Of course, this does not simply carry over to 
low-dimensional systems in condensed matter or
statistical physics:
Dimensionally hybrid action principles would not be a suitable 
tool for model building in theoretical investigations of these
systems because time derivatives generically will appear
only as bulk terms in the Lagrangian of a particle interacting
with a low-dimensional structure.

Therefore the main proposal of the present work is to use
{\it dimensionally hybrid Hamiltonians} in model building
for low-dimensional systems: A combination of two-dimensional
and three-dimensional terms and the ensuing interpolating 
correlation functions may help to narrow the gap between the
powerful methods of two-dimensional field theory and realistic
thin layers or surface structures in physics and technology.

In the sequel I will use this idea
to discuss non-relativistic particles interacting with
a thin layer. The system becomes a dimensionally hybrid system
with a specifically two-dimensional component
through the assumption that particles in the layer 
have a kinetic energy different from particles outside of the layer,
e.g.\ as a consequence of mass renormalization $M\to m$
due to the interaction of the particles with the
components of the layer. 
This yields some straightforward but interesting results.

In the next section I will show in a simple model that transmission
probabilities through thin layers in this class of models depend
also on the momentum parallel to the layer if $\mu=m(L)/L|_{L\to 0}$
remains finite. The discussion
of Green's functions in these models
will be the subject of Secs.\ \ref{sec:greens} and \ref{sec:greensK}.
Sec.\ \ref{sec:greens} contains in particular
a potential which interpolates between two-dimensional and
three-dimensional distance laws.

\section{Dimensionally hybrid Hamiltonians
and Green's functions}\label{sec:hybrid}

To investigate implications
of dimensionally hybrid Hamiltonians
for the description of the interaction of particles
with a thin layer we assume in the present section
that the layer is planar and homogeneous
and therefore
generates a potential $U(z)$, where $z$ is transverse to the layer.

In realistic two-dimensional
systems particles are not strictly bound to the layer,
and the effective particle mass in the layer may
be changed due to interactions.
This motivates the following Hamiltonian for particles
of mass $M$ in the presence of the layer 
\begin{eqnarray}\label{eq:HQFT}
H&=&
\left.
\frac{\hbar^2}{2\mu}\int\! d^2\vek{x}
\nabla\psi^+\cdot\nabla\psi\right|_{z=0}\\
\nonumber
&&
+\int\! d^2\vek{x}\int\! dz\left(\frac{\hbar^2}{2M}
\left(\nabla\psi^+\cdot\nabla\psi+\partial_z\psi^+
\cdot\partial_z\psi\right)+\psi^+U\psi\right).
\end{eqnarray}
Here and in the sequel all vectors are 2-dimensional
vectors in the layer.

The assumption behind the Hamiltonian (\ref{eq:HQFT})
 is that the same field $\psi$ may
describe e.g.\ free electrons in the bulk and large polarons\footnote{See e.g.
\cite{kittel,madelung,devreese} for introductions to polarons in solids.}
or other collective excitations involving conduction
electrons in the layer. 
The parameter $\mu$ has dimensions of mass per length, and in a limiting
procedure from layers of finite thickness $L$ would correspond to
\begin{eqnarray}\label{eq:mu}
\mu=\lim_{L\to 0}\frac{m(L)}{L},
\end{eqnarray}
where $m(L)$ would be the mass of the modes in the layer.

The corresponding 
equation of motion for stationary single-(quasi)particle
wavefunctions is
\begin{equation}\label{eq:schrodinger}
E\psi(\vek{x},z)=
-\delta(z)\frac{\hbar^2}{2\mu}\Delta\psi(\vek{x},0)
-\frac{\hbar^2}{2M}\left(\Delta+\partial_z^2\right)
\psi(\vek{x},z)+U(z)\psi(\vek{x},z).
\end{equation}

The Fourier {\it ansatz}
\begin{equation}\label{eq:fourier1}
\psi(\vek{x},z)=\frac{1}{2\pi}\int\!d^2\vek{k}\,\psi(\vek{k},z)
\exp(\mathrm{i}\vek{k}\cdot\vek{x})
\end{equation}
yields the separated equation
\begin{equation}\label{eq:sep1}
\left(E-\frac{\hbar^2k^2}{2M}\right)\psi(\vek{k},z)
=-\frac{\hbar^2}{2M}\partial_z^2\psi(\vek{k},z)
+U(z)\psi(\vek{k},z)+\delta(z)\frac{\hbar^2k^2}{2\mu}
\psi(\vek{k},0).
\end{equation}
Every solvable model of one-dimensional quantum mechanics
gives a solution to this class of layer models, with the kinetic
term of the layer modes only generating a cusp
proportional to $(M/\mu)k^2$ in $\ln\psi(\vek{k},z)$.

Obviously, the large longitudinal momentum modes are strongly
affected by the existence of layer modes, but we will see in a moment
that the two-dimensional kinetic term can also have
a strong impact on modes with small longitudinal momentum.\\

\begin{center}
\scalebox{0.8}{\includegraphics{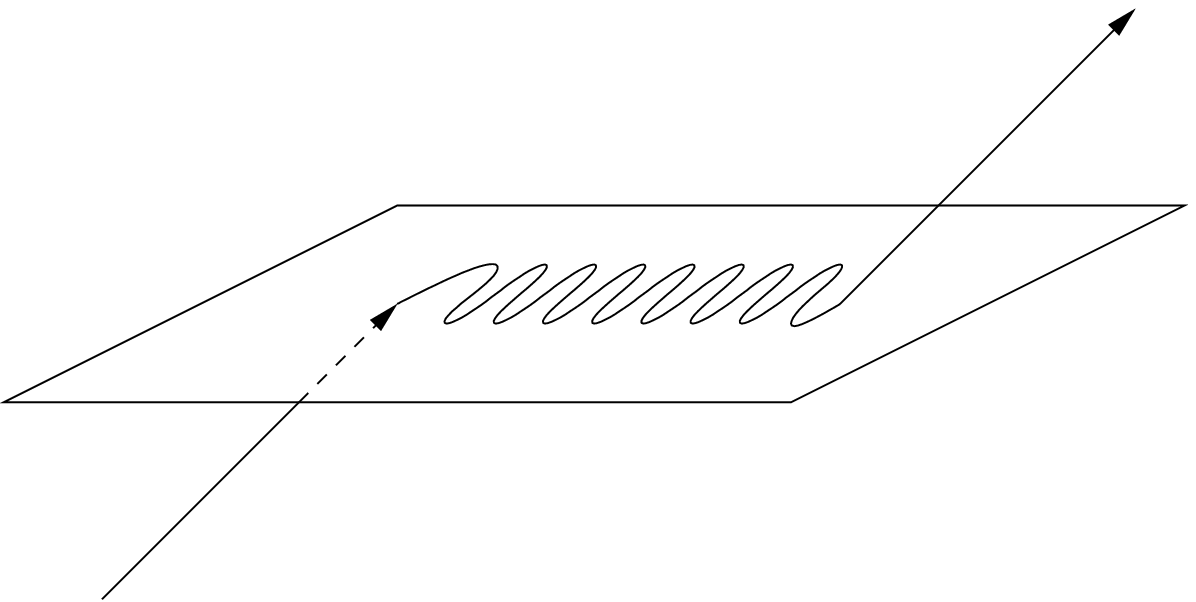}}
\end{center}

\noindent
Fig. 1: Contribution of a virtual planar mode to a particle
penetrating a thin homogeneous layer.\\[0.1ex]

It is a trivial exercise to adapt solutions of one-dimensional
quantum mechanics to the cusp imposed by the layer modes,
but it may be worthwhile to write down the modifications 
of the transmission coefficient due to the
layer modes when the layer potential represents
a work function: $U(z)=-w\delta(z)$.
The transmission coefficient for an infalling particle
of momentum $\{\vek{k},k_\perp\}$ is
\begin{equation}\label{eq:T}
T(\vek{k},k_\perp)=
\left[1+\frac{M^2}{k^2_\perp}\left(
\frac{w}{\hbar^2}-\frac{k^2}{2\mu}\right)^2
\right]^{-1},
\end{equation}
i.e.\ the layer modes increase the transmission probability
for low longitudinal momentum modes $0<\hbar^2k^2<4\mu w$,
and decrease the transmission probabilities for the modes of higher
longitudinal momentum.

This model also predicts a resonance in transmission for a
certain value
$\hbar^2k^2=2\mu w$ of the {\it longitudinal} momentum.
This is as a genuine consequence of the two-dimensional kinetic term
in (\ref{eq:HQFT}) and may be the simplest way to test
the viability of the idea of dimensionally hybrid Hamiltonians
in low-dimensional systems.

Obviously, the prediction of dependence of transmission probabilities
on longitudinal momenta requires finiteness of the parameter 
$\mu$ (\ref{eq:mu}), i.e.\ a derivation of the "phenomenological"
Hamiltonian (\ref{eq:HQFT}) from a limiting procedure of purely
three-dimensional models will require a thorough study of finite
size effects on mass renormalization in solids.

\section{Correlations on a layer}\label{sec:greens}

A model similar to (\ref{eq:HQFT}) allows for a neat discussion of the
impact of combinations of kinetic terms from different dimensions
on the "two-dimensional" correlation functions on the layer.

 For this we assume that the layer is not necessarily homogeneous,
but generates a strongly localized potential
\[
V(\vek{x},z)=u(\vek{x})\delta(z)
\]
along the layer:
\begin{eqnarray}\label{eq:HQFT2}
H&=&
\frac{\hbar^2}{2M}\int\! d^2\vek{x}\int\! dz
\left(\nabla\psi^+\cdot\nabla\psi+\partial_z\psi^+
\cdot\partial_z\psi\right)\\
\nonumber
&&
+\left.
\int\! d^2\vek{x}\left(\frac{\hbar^2}{2\mu}
\nabla\psi^+\cdot\nabla\psi
+\psi^+u\psi\right)\right|_{z=0}.
\end{eqnarray}

The generating functional for correlation functions on the layer
is\footnote{As usual
$\delta/\delta j$ acts from the right if $\psi$ is fermionic.}
\begin{equation}\label{eq:Z}
Z[j,j^+]=\int\! d\psi d\psi^+\exp\!\left(
-\beta H[\psi,\psi^+]
-\int\! d^2\vek{x}[\psi^+(\vek{x},0) j(\vek{x})+j^+(\vek{x})\psi(\vek{x},0)]
\right)
\end{equation}
\[
=
\exp\!\left(-\beta\int\! d^2\vek{x}
\frac{\delta}{\delta j(\vek{x})}u(\vek{x})
\frac{\delta}{\delta j^+(\vek{x})}\right)
Z_0[j,j^+]
\]
with
\begin{equation}\label{eq:Z0}
Z_0[j,j^+]\propto
\exp\!\left(\frac{2M}{\hbar^2\beta}
\int\! d^2\vek{x}\int\! d^2\vek{x}'\,j^+(\vek{x})G(\vek{x}-\vek{x}',0)
j(\vek{x}')\right).
\end{equation}
Since 
(\ref{eq:HQFT2}) is a free theory from a field theory point of view,
the "two-dimensional" correlations in it
can be calculated from tree-level diagrams, which involve
only the restriction
of the free Green's function $G(\vek{x}-\vek{x}',z)$ to the layer
and insertions of the layer potential.

The free Green's function used in (\ref{eq:Z0}) satisfies
\begin{equation}\label{eq:greeneq0}
\left(\Delta+\partial_z^2\right)G(\vek{x},z)
+2\ell\delta(z)\Delta G(\vek{x},0)=-\delta(\vek{x})\delta(z)
\end{equation}
where
\begin{equation}\label{eq:ell}
2\ell=\frac{M}{\mu}.
\end{equation}

The {\it ansatz}
\begin{equation}\label{eq:g0ans}
G(\vek{x},z)=\frac{1}{(2\pi)^3}\int\! d^2\vek{k}\int\! dk_\perp\,
G(\vek{k},k_\perp)\exp[\mathrm{i}(\vek{k}\cdot\vek{x}
+k_\perp z)]
\end{equation}
yields
\begin{equation}\label{eq:greeneqk0}
\left(k^2+k_\perp^2\right)G(\vek{k},k_\perp)
+\frac{\ell}{\pi}k^2 \int\! dk'_\perp\, G(\vek{k},k'_\perp)=1.
\end{equation}
This determines the $k_\perp$-dependence of the propagator
\begin{equation}\label{eq:kperpdep1}
G(\vek{k},k_\perp)=\frac{f(k)}{k^2+k_\perp^2}.
\end{equation}

With
\[
\frac{1}{\pi}\int_{-\infty}^\infty\! dk'_\perp\,
\frac{1}{k^2+k'^2_\perp}
=\frac{1}{k}
\]
we find the Green's function
\begin{equation}\label{eq:statickp}
G(\vek{k},k_\perp)
=\frac{1}{(1+k\ell)(k^2+k_\perp^2)}
\end{equation}
or
\begin{equation}\label{eq:staticz}
G(\vek{k},z)
=\frac{1}{2k(1+\ell k)}\exp(-k|z|).
\end{equation}

The solution\footnote{I follow the conventions of \cite{AS}
for Bessel and Struve functions.}
\begin{equation}\label{eq:pot}
G(\vek{x},z)
=\frac{1}{8\pi^2}\int_0^\infty\! dk\int_0^{2\pi}\!
d\varphi\,
\frac{\exp[k(\mathrm{i}r\cos\varphi-|z|)]}{1+k\ell}
=\frac{1}{4\pi}\int_0^\infty\! dk\,
\frac{\exp(-k|z|)}{1+k\ell}J_0(kr)
\end{equation}
of (\ref{eq:greeneq0})
can be thought of as the electrostatic potential of a unit charge
on the layer, if the fields which are continuous across the layer
make a special contribution to the Hamiltonian
of the electromagnetic field\footnote{In the spirit of the philosophy
advocated in this paper the electromagnetic fields in
(\ref{eq:Helm}) would simultaneously describe photons
in the bulk and polaritons in the layer.}:
\begin{equation}\label{eq:Helm}
H[F]=
\ell\int\! d^2\vek{x}\left(\vek{E}^2+B_\perp^2\right)
+\frac{1}{2}\int\! d^2\vek{x}\int\! dz
\left(\vek{E}^2+E_\perp^2+\vek{B}^2+B_\perp^2\right),
\end{equation}
e.g. as a consequence of a non-vanishing limit of
\[
2\ell=\lim_{L\to 0}(\epsilon_r L).
\]
Here $\epsilon_r$ is the relative permittivity of the layer
and $L$ its transverse extension.

 The perturbation
series (\ref{eq:Z},\ref{eq:Z0}) requires the Green's function
on the layer, which can be expressed as a linear combination
of a Struve function and a Bessel function of the second kind:
\begin{equation}\label{eq:Phi}
\Phi(\vek{x})=G(\vek{x},z)\Big|_{z=0}
=\frac{1}{8\ell}
\left[\mathbf{H}_0\!\left(\frac{r}{\ell}\right)
-Y_0\!\left(\frac{r}{\ell}\right)\right].
\end{equation}
This interpolates between two-dimensional and three-dimensional
distance laws
\[
r\ll\ell:\quad
\Phi(\vek{x})=\frac{1}{4\pi\ell}\left[
-\gamma-\ln\!\left(\frac{r}{2\ell}\right)
+\frac{r}{\ell}
+\mathcal{O}\!\left(\frac{r^2}{\ell^2}\right)\right],
\]
\[
r\gg\ell:\quad
\Phi(\vek{x})=\frac{1}{4\pi r}\left[
1-\frac{\ell^2}{r^2}+\mathcal{O}\!\left(\frac{\ell^4}{r^4}\right)
\right].
\]

$\Phi(\vek{x})$ along with the limiting cases is plotted in Fig.\ 2.\\[2mm]

\begin{minipage}[t]{70mm}
\noindent
Fig. 2: The solid line is the Green's function (\ref{eq:Phi})
on the layer
as a function of $x=r/\ell$, in units of $\ell^{-1}$. 
The upper dashed line is the 
three-dimensional $1/4\pi r$ potential, and the lower dashed line
is the two-dimensional logarithmic potential.
\end{minipage}
\hspace*{15mm}
\begin{minipage}[t]{6cm}
\vspace*{20mm}
\scalebox{0.45}{\includegraphics{223pot.eps}}
\end{minipage}

Polaron masses in semiconductors are of the order of effective
band masses. Due to strong band curvature the
ratios of polaron masses to the mass of free electrons 
can be of order $m/M\simeq 10^{-2}$, so one may hope that
in thin semiconducting layers of thickness $L$ the parameter
$\ell$ is of order $\ell\simeq 10^2L$.
Two-dimensional
distance laws for correlation functions might then be realized up to distances
of order $10L$, and intermittent behavior for distances between $10L$
and $200L$. For a possible realization of (\ref{eq:Phi}) as an electromagnetic
potential in thin layers, semiconducting compounds involving Pb might be the
best bet due to their high relative permittivities of order $\epsilon_r\sim 10^2
- 10^3$ \cite{LB}. The corresponding field strength on the layer is
\begin{equation}\label{eq:field}
-\partial_r\Phi(\vek{x})
=\frac{1}{8\ell^2}
\left[\mathbf{H}_1\!\left(\frac{r}{\ell}\right)
-Y_1\!\left(\frac{r}{\ell}\right)-\frac{2}{\pi}\right].
\end{equation}

\vspace*{4mm}

\begin{minipage}[t]{70mm}
\noindent
Fig. 3: The solid line is the field strength 
per charge (\ref{eq:field})
on the layer
as a function of $x=r/\ell$, in units of $\ell^{-2}$. 
The dashed line approaching the solid line for $x<1$ is the 
two-dimensional $1/4\pi\ell r$ field, and the other dashed line
is the three-dimensional $1/4\pi r^2$ field.
\end{minipage}
\hspace*{15mm}
\begin{minipage}[t]{6cm}
\vspace*{20mm}
\scalebox{0.45}{\includegraphics{223force.eps}}
\end{minipage}

This reduction of the force between charges
in a thin dielectric layer with finite
$2\ell=\lim_{L\to 0}(\epsilon_r L)$ can pictorially be understood as
a consequence of the fact that field lines are refracted away
from the layer when they leave the layer. This reduces the field lines e.g.\ 
between two opposite charges at short distances, since the field
lines cannot re-enter the
layer on short scales, whereas for large separation of the
two charges along the layer re-entry renders the refraction effect
negligible.

\section{The Green's function for scattering
from two-dimensional potentials on the layer}\label{sec:greensK}

The stationary wave equation from (\ref{eq:HQFT2}) is
\begin{equation}\label{eq:schrodinger2}
E\psi(\vek{x},z)=
\delta(z)\left(-\frac{\hbar^2}{2\mu}\Delta+u(\vek{x})\right)\psi(\vek{x},0)
-\frac{\hbar^2}{2M}\left(\Delta+\partial_z^2\right)
\psi(\vek{x},z).
\end{equation}

We have already noticed that this can be solved exactly for
$u(\vek{x})$=constant.

In discussing scattering of bulk particles from the layer in the
model (\ref{eq:HQFT2}) we could proceed using the ordinary three-dimensional
Green's function for scattering of waves of energy $E=\hbar^2K^2/2M$
and treat the full two-dimensional contribution to Eq.\
(\ref{eq:schrodinger2}) as a perturbation. However, 
here I rather would like
to treat only the layer potential $u(\vek{x})$
as a perturbation. This has the virtue of reducing
the perturbation for large longitudinal momenta.

The relevant unperturbed wave is then
\begin{eqnarray}\label{eq:psiin}\nonumber
\psi_0(\vek{x},z)&=&\frac{1}{\sqrt{2\pi}^3}
\exp\!\left(\mathrm{i}\vek{K}_\|\cdot\vek{x}\right)
\left[\Theta(-z)\left(\exp\!\left(\mathrm{i}K_\perp z\right)
+\frac{K_\|^2\ell}{\mathrm{i}K_\perp-K_\|^2\ell}
\exp\!\left(-\mathrm{i}K_\perp z\right)\right)\right.\\
&&\left.
+\Theta(z)\frac{K_\perp}{K_\perp+\mathrm{i}K_\|^2\ell}
\exp\!\left(\mathrm{i}K_\perp z\right)
\right]
\end{eqnarray}
where again the definition (\ref{eq:ell}) was used.

The relevant Green's function $G_K$
for propagation of bulk plane waves of energy
\[
E=\frac{\hbar^2 K^2}{2M}=\frac{\hbar^2}{2M}\left(\vek{K}_\|^2+K_\perp^2\right)
\]
 has to satisfy
\begin{equation}\label{eq:greeneq}
\left(\Delta+\partial_z^2+ K^2\right)G_K(\vek{x},z)
+2\ell\delta(z)\Delta G_K(\vek{x},0)=-\delta(\vek{x})\delta(z),
\end{equation}
and the solution proceeds similarly to the solution of (\ref{eq:greeneq0}).
The Fourier {\it ansatz} (\ref{eq:g0ans}) yields
\begin{equation}\label{eq:greeneqk}
\left(k^2+k_\perp^2- K^2\right)G_K(\vek{k},k_\perp)
+\frac{\ell}{\pi}k^2 \int\! dk'_\perp\, G_K(\vek{k},k'_\perp)=1,
\end{equation}
which determines the $k_\perp$-dependence of the propagator
\begin{equation}\label{eq:kperpdepk1}
G_K(\vek{k},k_\perp)=\frac{f(k)}{k^2+k_\perp^2- K^2}.
\end{equation}
At this stage the possibility of poles complicates the calculation slightly:
In evaluating the integral in (\ref{eq:greeneqk})
with (\ref{eq:kperpdepk1}) for $k<K$
we have to make a judicious choice on how to shift the poles
or the integration path
at $k_\perp=\pm\sqrt{K^2-k^2}$, corresponding to 
 correct physical boundary conditions on $G_K(\vek{k},z)$.
The correct choice turns out to be
$k_\perp=\pm(\sqrt{K^2-k^2}+\mathrm{i}\epsilon)$
since $G_K(\vek{k},z)$ is supposed to describe outgoing scattered waves
from the layer if $k<K$, i.e.\ we have
\begin{eqnarray}\label{eq:kperpdep2}
G_K(\vek{k},k_\perp)&=&\frac{f(k)}{k^2+k_\perp^2-K^2-\mathrm{i}\epsilon}\\
\nonumber
&=&f(k)\left(\mathcal{P}\frac{1}{k^2+k_\perp^2-K^2}
+\mathrm{i}\pi\delta(k^2+k_\perp^2-K^2)\right).
\end{eqnarray}
With
\[
\frac{1}{\pi}\int_{-\infty}^\infty\! dk'_\perp\,
\frac{1}{k^2+k'^2_\perp-K^2-\mathrm{i}\epsilon}
=\frac{\Theta(k^2-K^2)}{\sqrt{k^2-K^2}}
+\mathrm{i}\frac{\Theta(K^2-k^2)}{\sqrt{K^2-k^2}}
\]
we find the Green's function
at large longitudinal wavelength $k<K$ 
\begin{equation}\label{eq:greenlowk}
G_K(\vek{k},k_\perp)
=\frac{\sqrt{K^2-k^2}}{(\sqrt{K^2-k^2}+\mathrm{i}k^2\ell)
(k^2+k_\perp^2-K^2-\mathrm{i}\epsilon)},
\end{equation}
\begin{equation}\label{eq:greenlowk2}
G_K(\vek{k},z)
=\frac{1}{2k^2\ell-2\mathrm{i}\sqrt{K^2-k^2}}
\exp\!\left(\mathrm{i}\sqrt{K^2-k^2}|z|\right),
\end{equation}
while the short longitudinal wavelength part is 
\begin{equation}\label{eq:greenlargek}
G_K(\vek{k},k_\perp)
=\frac{\sqrt{k^2-K^2}}{(\sqrt{k^2-K^2}+k^2\ell)
(k^2+k_\perp^2-K^2-\mathrm{i}\epsilon)},
\end{equation}
\begin{equation}\label{eq:greenlargek2}
G_K(\vek{k},z)
=\frac{1}{2\sqrt{k^2-K^2}+2k^2\ell}\exp\!\left(-\sqrt{k^2-K^2}|z|\right).
\end{equation}
Of course, the Green's function again reduces to the usual
three-dimensional
result for $\ell\to 0$, i.e.\ if the modes in the layer become so heavy
that they decouple. 

With an incoming plane wave the
integral equation following from (\ref{eq:schrodinger2},\ref{eq:greeneq})
is
\begin{equation}\label{eq:inteq}
\psi(\vek{x},z)
=\psi_0(\vek{x},z)
-\frac{2M}{\hbar^2}
\int\! d^2\vek{x}'\,G_K(\vek{x}-\vek{x}',z)u(\vek{x}')\psi(\vek{x}',0).
\end{equation}
In a Born approximation this yields with (\ref{eq:psiin})
\begin{equation}\label{eq:psiborn}
\psi(\vek{x},z)
=\psi_0(\vek{x},z)
-\frac{2M}{\sqrt{2\pi}^7\hbar^2}\frac{K_\perp}{K_\perp+\mathrm{i}K_\|^2\ell}
\int\! d^2\vek{k}\,\exp(\mathrm{i}\vek{k}\cdot\vek{x})
G_K(\vek{k},z)u(\vek{k}-\vek{K}_\|),
\end{equation}
where the normalization of the Fourier transformed layer potential is
\[
u(\vek{q})=\int\! d^2\vek{x}\,\exp(-\mathrm{i}\vek{q}\cdot\vek{x})u(\vek{x}).
\]

Eq.\ (\ref{eq:psiborn}) together with (\ref{eq:greenlargek2}) implies that
no particular scattering wave is generated by the short wavelength components
 at $|\vek{q}|<K_\|+\sqrt{K_\|^2+K_\perp^2}$
of the layer potential, which is simply a statement of the limited resolving
power of the external wave. However, beyond that the $\ell$-dependence of
(\ref{eq:greenlowk2}) and (\ref{eq:psiborn}) implies that the two-dimensional
kinetic term in (\ref{eq:HQFT2}) reduces potential scattering at large $K_\|$.


\section{Summary}\label{sec:conc}

Two-dimensional field theory is a very appealing subject with many powerful
results. One reason for this is because every tensor and spinor field on a 
2-manifold decomposes into
covariant primary fields which provide 1-dimensional representations
of the corresponding symmetry groups\footnote{This is true
beyond the realm of conformal transformations if the Beltrami
parameters on the 2-manifold are used to decompose tensors
and spinors into 
{\it covariant
primary fields}, see Secs.\ 1 and 2 in \cite{rdfp1} for tensors
and \cite{hermann} for spinors.}.

However, the assumption of strict two-dimensionality seems too restrictive
when it comes to comparisons with actual layers or surface structures in
physics.
Conservative theoretical approaches to low-dimensional structures in 
physics and technology therefore rely on 
genuine three-dimensional Hamiltonians and only restrict the 
locations and momenta of particles to a surface or a layer 
(see e.g.\ Sec.\ 9.2 in \cite{madelung}). In these approaches
two-dimensionality is only taken into account at a kinematical level,
at the expense of sacrificing the powerful methods and results of
two-dimensional field theory.

On the other hand, recent results in brane theory taught us that
straightforward combinations of four-dimensional terms and five-dimensional
terms in action principles yield interpolating Green's functions
on the brane,
and it is apparent from the functional integral representation that
this property must also hold for higher order correlation functions on the
brane.

As mentioned above this cannot carry over directly to low-dimensional systems
in condensed matter and statistical
physics, but it initiated the present proposal to use dimensionally
hybrid Hamiltonians $H=\ell h_2+H_3$
for theoretical investigations of low-dimensional
structures in physics.

Such an approach has the prospect to provide more realistic results
 than strictly two-dimensional field theory, while at the same time
utilizing the power of two-dimensional field theory for the
determination e.g.\ of equilibrium correlation
functions in the limiting cases $\ell\to\infty$ or $k\ell\gg 1$.

 Further virtues of this approach are predictions on the
transition behavior between two-dimensional and three-dimensional
distance laws in layers and surface structures, and a better understanding
on how two-dimensional structures might be approached in the more
conservative purely three-dimensional framework, through studies
of finite size effects on effective masses and permittivities
in three-dimensional models.

To illustrate the use and some straightforward consequences
of dimensionally hybrid
Hamiltonians a homogeneous layer
and layers with strongly localized potentials 
were studied. In these settings the
 field $\psi$ describes simultaneously bulk particles of mass $M$
and excitations of mass $\mu$ per transverse length in the layers.
These models 
might serve as an approximation to thin layers of semiconductors
or polar solids, where the planar modes would correspond to large
polarons.
The model predicts a strong dependence of transmission probabilities
on longitudinal momentum.

The Green's function in the model also yields
a static potential which interpolates between the logarithmic
two-dimensional distance law at distances $\ll\ell$ and the three-dimensional
Coulomb law at distances $\gg\ell$.\\[1ex]


\noindent
{\bf Acknowledgement:} This work was supported by NSERC.

\end{document}